# The Computational Mechanisms of Detached Mindfulness

**Brendan Conway-Smith (brendan.conwaysmith@carleton.ca),**
**Robert L. West (robert.west@carleton.ca)**
Department of Cognitive Science, Carleton University, Ottawa, ON K1S 5B6 Canada

**Abstract**

This paper investigates the computational mechanisms underlying a type of metacognitive monitoring known as detached mindfulness, a particularly effective therapeutic technique within cognitive psychology. While research strongly supports the capacity of detached mindfulness to reduce depression and anxiety, its cognitive and computational underpinnings remain largely unexplained. We employ a computational model of metacognitive skill to articulate the mechanisms through which a detached perception of affect reduces emotional reactivity.

**Keywords**: metacognition; mindfulness; affect; emotion; proceduralization; ACT-R; Common Model

## Introduction

The attempt to build a Unified Cognitive Architecture (Newell, 1994) that can replicate human-like intelligence must necessarily account for the routine interplay between affect and metacognitive processes. Historically, cognitive modeling research has focused predominantly on knowledge-based processing such as reasoning, vision, and AI problem-solving, with little or no computational account of the critical role of emotion and metacognition.

This need for increased computational understanding is underscored by the fact that perseverative patterns of negative emotion, such as depression and anxiety, are the largest causes of cognitive disability worldwide (World Health Organization, 2022). Consequently, there has been a global push to develop metacognitive techniques that allow individuals to engage with their emotions adaptively. A particularly effective metacognitive technique is referred to as 'detached mindfulness' (Wells, 2005). This technique focuses on developing one's perception of the momentary changes of affective states, shown to significantly reduce feelings of distress, emotional reactivity, and to improve overall cognitive functioning (Hammersmark et al., 2024).

While decades of clinical research strongly supports the effectiveness of metacognitive strategies and detached mindfulness in particular, their underlying cognitive and computational mechanisms remain largely unexplained. This paper will investigate the cognitive and computational constituents that underpin detached mindfulness and its therapeutic benefits. Specifically, we will discuss the metacognitive mechanisms by which the perception of affective fluctuations deactivates emotional reactivity.

For this purpose, we will employ the Common Model of Cognition (CMC), originally the 'Standard Model' (Laird, Lebiere, & Rosenbloom, 2017), which provides a unified framework for investigating the fundamental elements of cognitive and metacognitive phenomena. By utilizing the Common Model, and specifically ACT-R (Anderson & Lebiere, 1998) in this investigation, we intend to address important questions largely unexplored in cognitive models: How does metacognitive training in detached mindfulness reduce persevering styles of negative emotions? By what computational mechanism does perceiving the momentary changes in affect disengage emotional reactivity such as meta-emotions?

First, we will overview the relevant literature on metacognition and mindfulness techniques. Second, we outline the computational mechanisms involved in a model of metacognitive skill learning. Third, we apply this model of metacognitive skill learning to detached mindfulness to clarify its underling components and the precise mechanism by which it reduces emotional reactivity as reported in the literature.

## Metacognition

We propose that an active mechanism of detached mindfulness fundamentally relies on a form of automatized metacognition. The common conception of metacognition refers to the monitoring and control of cognitive processes (Flavell 1979; Fleming, Dolan, & Frith, 2012). *Metacognitive control* refers to the active regulation of cognitive processes or states to either activate or inhibit them (Proust, 2013; Wells, 2019). The regulation of one's own cognitive processes can involve various processes such as attention, emotion, planning, reasoning, and memory (Efklides, Schwartz, & Brown, 2017; Pearman et al., 2020). *Metacognitive monitoring* refers to the capacity to recognize and identify cognitive states. It involves the perception of internal mental states such as thoughts and feelings in order to regulate those states or direct behavior.

Studies demonstrate that metacognitive monitoring can be developed and improved through training (Baird, Mrazek, Phillips, & Schooler, 2014). For instance, attentional processes can be developed and enhanced through the repeated practice of attention-based tasks (Posner et al., 2015). Metacognitive training such as mindfulness techniques is integral to both Cognitive Behavior Therapy (CBT; Dobson, 2013) and

Metacognitive Therapy (MCT; Normann & Morina, 2018) and facilitates improved control over maladaptive thoughts and emotions (Wells, 2011, 2019; Hagen et al., 2017). The benefits of mindfulness training rely partly on its enhancement of metacognitive sensitivity, which is the extent to which one is able to perceive their own mental processes or states, including thoughts, feelings, and emotions (Fleming & Lau, 2014). Improved metacognitive sensitivity has the effect of lowering one's metacognitive threshold — the minimal level of a stimulus required for a person to be aware of some mental state and make a judgment about it (Charles, Chardin, & Haggard, 2020; Pauen & Haynes, 2021). The metacognitive threshold can also be lowered by way of attentional training, such as detached mindfulness and meditation, which allows one to perceive a weaker signal strength from internal cognitive states (Fox et al., 2016). While this has been effectively modelled within ACT-R (Conway-Smith & West, 2023) it is not the main focus of this paper.

## Metacognition as mindfulness

Scientific interest in mindfulness practice has become a target of interdisciplinary research and has grown exponentially over the past few decades (Van Dam et al., 2018). Metacognition and mindfulness are often used interchangeably within cognitive psychology (Holas & Jankowski, 2013). Mindfulness psychology contends that a significant degree of emotional distress and pathological symptoms are caused by the illusory perception of affective experience being more permanent than it actually is. This perceptual illusion has been explained as the result of poor metacognitive sensitivity that obscures the detection of affective fluctuation (Brown & Ryan, 2003; Grossman et al., 2010). To address this metacognitive deficiency, detached mindfulness has emerged as a uniquely effective therapeutic technique (Wells & Matthews, 1994; Hammersmark et al., 2024). This involves participants learning to observe moment-to-moment changes in mental states, including subtle emotional fluctuations, and allowing these states to occur without engaging with or reacting to them.

This non-reactive state of awareness is also referred to as 'equanimity'. In mindfulness therapies that do not promote equanimity, awareness alone is often insufficient to increase subjects' psychological well-being (Cardaciotto et al., 2008). Detached mindfulness is most closely aligned with Vipassana meditation (in the tradition of S.N. Goenka), an old and popular technique that largely focuses on cultivating equanimity i.e., perceptual sensitivity to variations in affect and physical sensation (Kakumanu et al., 2018). Regular practice of this technique has shown to improve executive functioning, response inhibition, and control over emotional reactions such as meta-emotions (Andreu et al., 2019).

## Meta-emotion

Meta-emotions are emotions that automatically react to other emotions (Jäger & Banninger-Huber, 2015; Predatu, David & Maffei, 2020). For instance, a primary negative emotion (sadness) can cause a greater secondary negative emotion (despair) which may cause an even greater tertiary negative emotion (depression). Meta-emotions are instances of positive feedback, in which an emotional response to a primary emotion intensifies the overall emotional experience, leading to an amplified response. Meta-emotions occur as low-level reactive processes that are largely unconscious and involuntary, making them difficult to intervene in.

While therapeutic practices aim to control the resulting effects of meta-emotions such as anxiety and depression, techniques such as detached mindfulness and Vipassana aim to address the source, which is considered the false perception of affective permanence. To date, we lack a mechanistic understanding of precisely how detached mindfulness breaks through the illusion of affective permanence and disengages emotional reactivity. To clarify this mechanism, we will apply a model of metacognitive skill that articulates the components involved in this process and how they interact. Central to this explanation is a process referred to as proceduralization, a framework common among skill theories. We will first discuss the relevant components of metacognition and their expression in the cognitive architecture ACT-R. We will then explore how the components of proceduralization function to produce the therapeutic mechanism active in detached mindfulness.

## Components of metacognition

There are at least two types of cognitive representations that can engage in metacognitive monitoring and control processes — declarative knowledge and procedural knowledge. Metacognitive knowledge, or meta-knowledge, is considered a form of declarative knowledge (Schraw & Moshman, 1995; McCormick, 2003; Wells, 2019). Meta-knowledge takes the form of an explicit metarepresentation that is propositionally formatted and refers to a cognitive property, e.g., "I am focused" (Shea et al., 2014; Proust, 2013). Meta-knowledge can also take the form of a metacognitive instruction, which specifies a mental action to be performed (Wells, 2019). A metacognitive instruction, or meta-instruction, prescribes an action directed toward controlling some cognitive process, e.g., "Focus on the current task." Metacognitive knowledge is considered to be distinct from metacognitive skill, as it does not automatically lead to the deployment of metacognitive processes (Veenman & Elshout, 1999).

The execution of metacognitive instructions is performed by way of procedural knowledge. Improvements in metacognition are said to involve the refining of procedural knowledge that people use to monitor and control their own cognitive processes (Brown & DeLoache, 1978; Schraw & Moshman, 1995; Wells, 2019). The various realms of metacognitive skills can be understood as different domains of procedural knowledge (Veenman et al., 2005).

## ACT-R

Various theories of metacognition have been modelled within the ACT-R cognitive architecture (Reitter, 2010; Anderson & Fincham, 2014). ACT-R instantiates decades of research on how human cognition functions computationally. Its mandate is to depict the components necessary for human intelligence, which include working memory, perception, action, declarative memory, and procedural memory. These modules have also been correlated with their associated brain regions (Borst et al., 2015).

The ACT-R cognitive architecture fundamentally distinguishes between declarative and procedural knowledge, which accords with the literature on skill acquisition in philosophy and psychology (Squire, 1992; Christensen, Sutton, & McIlwain, 2016). Declarative knowledge is formatted propositionally and structured within semantic networks. Procedural knowledge is commonly referred to by researchers as containing "procedural representations" (Anderson, 1982; Pavese, 2019). Within ACT-R, procedural representations are computationally specified as "production rules" which are a dominant form of representation within accounts of skill (Newell, 1994; Taatgen & Lee, 2003; Anderson et al., 2019). Neurologically, production rules are associated with the 50ms decision timing in the basal ganglia (Stocco, 2018). Production rules, or "productions", transform information and change the state of the system to complete a task or resolve a problem. A production rule is modeled after a computer program instruction in the form of a "condition-action" pairing. It specifies a condition that, when met, performs a prescribed action. A production is also thought of as an "if-then" rule. *If* the condition is satisfied, such as matching to working memory, *then* it fires an action (Figure 1).

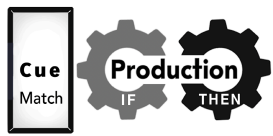

Figure 1: Production rules are formatted as a condition-action pairing. IF the condition side matches to the cue in working memory, THEN it fires an action.

Affect have been modelled computationally within ACT-R as non-propositional representations in working memory, or "metadata" (West & Conway-Smith, 2019). These types of affective information, encompassing both emotional states and noetic feelings, are essentially regarded as patterns within working memory that can be accessed by production rules.

Production rules match to and fire off the content in working memory. Should any stimuli or pattern appear in working memory, productions that match this pattern will arise from procedural memory and fire a prescribed action. In this way, cues in working memory can prompt procedural knowledge to act within various domains — motor, cognitive, and metacognitive. It is these specific cognitive units that are developed and refined during the process of proceduralization.

## Proceduralization

The concept of proceduralization is often used within the skill acquisition literature to explain the cognitive mechanisms involved in task learning (Fitts & Posner, 1967; Dreyfus & Dreyfus, 1986; Kim & Ritter, 2015). It refers to the process by which a task becomes automated, allowing it to be performed more efficiently and accurately, with minimal conscious effort or attention. The process involves converting slow declarative knowledge into fast procedural knowledge which is then increasingly refined. Skill performance can be further improved by way of mechanisms such as time delayed learning, where faster productions are rewarded. Proceduralization plays a significant role in the cognitive processes involved in skill learning within domains such as motor skill, cognitive skill, and metacognitive skill (Fitts, 1964; Anderson, 1982).

### Metacognitive proceduralization

Metacognitive proceduralization involves a mechanism by which human cognition becomes more skillful at monitoring and controlling its own processes, such as attention, emotion, and metacognitive sensitivity (Conway-Smith, West, & Mylopoulos, 2023). Previous research has presented proceduralization as a mechanism that can lower the metacognitive threshold, allowing one to perceive increasingly weaker signals from mental states and more subtle changes in affect (Conway-Smith & West, 2023). It is hypothesized that proceduralization accomplishes this through the building and refining of simpler, faster production rules. Faster and less complex productions, particularly those that notice internal states, increase the chances of picking up fleeting or intermittent signals related to emotions and epistemic feelings, such as confidence and feelings of knowing (FoK). However, this model does not address the process by which it mitigates emotional reactivity. By extending this research on metacognitive proceduralization, we can investigate a mechanism whereby sufficient metacognitive sensitivity can be developed to deactivate meta-emotions.

Metacognitive skill progresses through stages that parallel those of motor skill and cognitive skill, from an early stage of instruction-following to an expert stage that relies on automatic procedural knowledge (production rules).

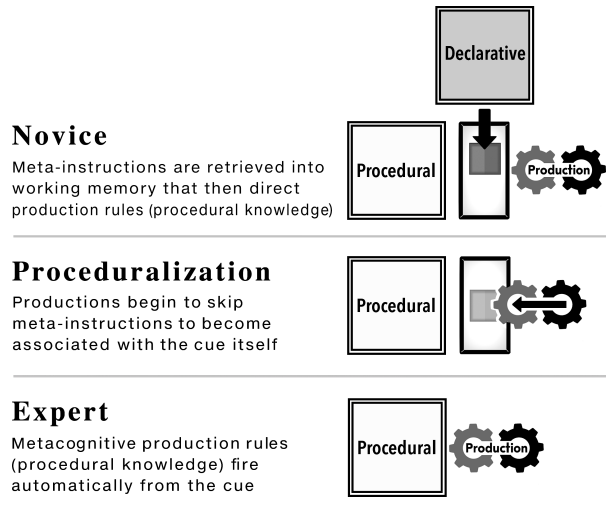

Figure 2: Three stages of metacognitive skill learning through the process of proceduralization (Conway-Smith, West, & Mylopoulos, 2023).

Metacognitive training in detached mindfulness progresses through the following three stages (Figure 2):

**The novice stage** involves the use of written or verbal meta-instructions to monitor or control some cognitive state (such as attentional training or meditation). In the case of metacognitive training in equanimity, meta-instructions direct the novice's attention toward the momentary changes of affective experience (feeling, sensation, or emotion). These meta-instructions are carried out by productions that retrieve them from declarative memory and execute them. Initial metacognitive performance is slow, effortful, error-prone, and requiring a large degree of working memory.

**The intermediate stage** of metacognitive training involves proceduralization, where practicing meta-instructions leads to the creation of faster production rules for accomplishing tasks. Specifically, repeated practice results in the compilation of task-specific production rules that bypass declarative knowledge. Because these rules are faster — due to bypassing declarative memory and possibly being less complex — they are more strongly rewarded and more likely to bypass the retrieval of instructions in the future. As a result, metacognitive performance becomes quicker, less effortful, and more automatic.

**The expert stage** involves a robust accumulation of production rules that have been refined and stored in procedural memory. These productions can be deployed automatically to act out monitoring and control processes quickly and effectively. These productions may be faster and less complex, resulting in a lower metacognitive threshold and an improved perception of affective experience. Metacognitive performance in this case demonstrates many characteristics of expertise, i.e., being fast, effective, automatic, and requiring minimal working memory.

## Deactivating meta-emotions

Proceduralization, the development of task-specific production rules, assists in providing a computational account of how training to perceive affective variations (equanimity) results in the deactivation of meta-emotions.

Recall that production rules match and fire off the content of working memory at a default rate of 50ms. That is, productions require at least 50ms to detect a pattern held within working memory. Should a pattern be perceived as sufficiently stable for over 50ms, productions will automatically match and fire off that pattern. Hence, the timing of production rules may be considered a condition of the metacognitive threshold (and psychophysical thresholds more generally) as it provides a partial account of which properties of the stimulus are needed to evoke a response, i.e., strength of signal and perceived stability.

An analogous psychophysical threshold is well known in vision research, where a light that flickers rapidly enough appears to be constant (Landis, 1954). This visual illusion is exploited in film production, where still frames are sped up to 24 frames per second to give images the appearance of consistency. The visual threshold at which still images appear to be constant has been referred to as the "moment of fusion". This visual threshold can be partially raised or lowered due to individual differences such as fatigue and age. For our purposes, the illusion of the flicker-fusion phenomena is comparable to the illusion of affective stability, in that they both rely on a person's inability to perceive change above a certain rate.

Similar to the visual threshold, an individual's metacognitive threshold is variable and can be lowered through attention training to perceive weaker signals from internal cognitive states, such as subtle changes in affect. Proceduralization offers a mechanism for developing and refining production rules that are more sensitive to internal signals, so as to eventually break the illusion of affective consistency.

A key insight into precisely how the refined perception of affective change (equanimity) deactivates emotional reactivity comes from the timing of production rules.

## Above the threshold

To the extent that a person's metacognitive threshold is above the 50ms firing rate of production rules, they will perceive any pattern within working memory to be relatively stable. Should a negative emotion appear to be consistent over the 50ms threshold, productions have sufficient time to match and fire a secondary negative emotion in response to the first. Assuming the same conditions, the secondary negative emotion may be perceived and reacted to again, producing a tertiary negative emotion. As long as the metacognitive threshold remains, along with the illusion of affective consistency, production rules may fire automatically, and emotional reactivity may repeat indefinitely.

This explanation sheds light on a potential mechanism that generates the continuous increase in negative emotions as experienced within many psychological disorders. Increasing and persistent cycles of maladaptive emotions are among the most common symptoms of mental illnesses and are associated with Cognitive Attentional Syndrome (CAS; Wells, 2009). A nearly universal phenomenon in cognitive disorders, CAS is a style of negative processing marked by fixed, negatively-biased attention which causes maladaptive emotions to be preserved and heightened, resulting in a continual state of emotional distress.

While there is a lack of computational explanations for the mechanisms underlying this style of maladaptive processing, the timing of production rules can help explain how negatively valenced emotions can be heightened through a process of positive feedback. Production rules also help explain the largely unconscious and involuntary nature of emotional reactions, underscoring the need for metacognitive training to develop productions that counteract them.

## Below the threshold

We propose that a key mechanism contributing to the deactivation of meta-emotions is the ability to perceive affective change below the 50ms firing rate of production rules. Reducing the metacognitive threshold below 50ms produces an effect similar to the visual flicker-fusion illusion that occurs when the film speed is reduced below 24 frames per second. The illusion of consistency is broken, and one perceives the rapid arising and passing of experience.

This refined perception of affective variations inhibits production rules from matching to the constant fluctuations in working memory (Figure 3). In effect, production rules do not have enough time to identify the rapidly changing pattern of affect. In principle, as long as sufficient metacognitive sensitivity remains, productions are unable to fire secondary emotions.

Lowering one's metacognitive threshold below the 50ms rate requires an expert level of metacognitive skill, as it necessitates the accumulation of sufficiently refined production rules. These expert production rules are better able to detect subtle variations in affective experience and fleeting signals from other internal cognitive states. Conversely, if one's metacognitive threshold rises above 50ms, the affective pattern may appear stable enough for emotional reactivity to resume.

This account helps articulate how the subcomponents of mindfulness training assist in diminishing cycles of negative emotion in psychological disorders such as Cognitive Attentional Syndrome. Individuals who experience CAS are often caught in patterns of negative emotion without a normal exit condition from the informational loop (Wells, 2019). From a computational standpoint, the development of production rules of the type discussed would provide an exit condition from maladaptive emotional loops that would otherwise persist.

This analysis highlights the pivotal role of metacognitive training in emotional regulation and the key mechanism by which metacognitive practices such as detached mindfulness enhance the ability to perceive emotions without reacting to them.

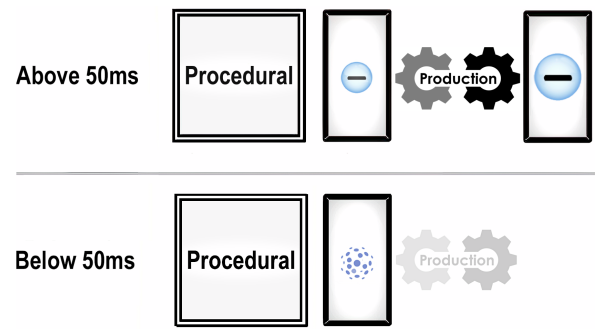

Figure 3. Above the 50ms threshold, an emotion is perceived as sufficiently stable for productions to match and fire secondary emotions. Below the 50ms threshold, the perception of emotional impermanence prevents productions from matching and firing secondary emotions.

## Other considerations

Accounting for mindfulness with cognitive modeling is a multifaceted endeavour, and there are many other considerations. For example, there is the issue of buffer decay, or how long patterns of activity can remain within working memory. These issues would apply to representations of both thought and emotion. Another issue is the ability for productions to match to emotional states and to declaratively label them. A particular issue that arises here can be understood in terms of partial matching, or the fidelity of the match. If we take emotion to be a representation of neural activity then we would expect it to have gradations of variability. Since the ability to recognize emotions would depend on our ability to match to these representational gradients, we would need to assume some form of fuzzy matching. This raises the possibility that some individuals could have more finely tuned productions and conceptual categories for matching emotions, while others may have broader, more fuzzy categories.

Finally, Conway-Smith and West (2023) argued that the capacity of production rules to speed up could increase one's sensitivity to detecting shifts in emotion, and discussed various ways that this speed up could be modeled.

## Conclusion

In this paper we have argued that Common Model type architectures can account for important aspects of mindfulness and meditation practices. In particular, we have employed the concept of metacognitive proceduralization to explore the mechanism by which detached mindfulness disengages meta-emotions. A complete model has yet to be constructed, as more theoretical work is required to determine a method of evaluation, considering there is presently no obvious data source with which to compare. One future possibility would be to better articulate the neural correlates of this model and to compare these to the neural imaging results of meditators.

By elucidating the computational processes involved in detached mindfulness and its influence on emotional reactivity, we contribute to a more comprehensive computational understanding that integrates both metacognitive monitoring and control within a unified framework. Meditation on the impermanence of affect is presently an edge case for the Common Model, one that will likely raise questions as to its capacity to simulate it. Our analysis demonstrates that the Common Model framework is able to interpret this practice in a way that accords with reports from practitioners, i.e., the stages of learning, their experiences, and their ability to apply it.

Moreover, by applying the ACT-R cognitive architecture to the study of metacognitive proceduralization, we help bridge the gap between cognitive modeling and psychological practice. The exploration of metacognitive proceduralization within the framework of the Common Model, and specifically ACT-R, offers a novel approach to understanding and intervening in the cycle of negative emotional reactions. Our approach facilitates the exploration of previously underexamined facets of cognitive modeling, aiding in the development of a more complete and integrated cognitive architecture.